\documentclass{elsart}
\usepackage{graphics,natbib,graphicx,psfig,amssymb} 
\journal{New Astronomy}
\input psfig.sty
\begin{document}
\begin{frontmatter}
\title{The metallicity distributions in high-latitudes with SDSS}
\author[istanbul]{S. Ak\corauthref{cor}},
\corauth[cor]{corresponding author.}
\ead{akserap@istanbul.edu.tr}
\author[istanbul]{S. Bilir},
\author[beykent]{S. Karaali},
\author[basel]{R. Buser}
\author[Spain1,Spain2]{and A. Cabrera-Lavers}
\address[istanbul]{Istanbul University, Faculty of Sciences, Department of Astronomy and Space Sciences, 34119 University, Istanbul, Turkey}
\address[beykent]{Beykent University, Faculty of Science and Letters, Department of Mathematics and Computing, 
          Beykent 34398, Istanbul, Turkey}
\address[basel]{Astronomisches Institut der Universit\"{a}t Basel, Venusstrasse 7, 4102 Binningen-Switzerland}
\address[Spain1]{Instituto de Astrof\'isica de Canarias, C/V\'ia L\'actea, s/n, 38200 La Laguna, Tenerife, Spain}
\address[Spain2]{GTC Project Office, C/V\'ia L\'actea s/n, 38200, La Laguna, Tenerife, Spain}

\begin{abstract}
We present metallicities and their variations with different parameters for 36 high-latitude fields covering Galactic longitudes $0^\circ < l \leq360^\circ$. The metallicities for relatively short vertical distances ($z<2.5$ kpc) show systematic fluctuations with Galactic longitude, similar to those of the thick-disc scaleheight, which may be interpreted as indicating a common origin, viz., the flare effect of the disc (\citealt{Bilir07}). This suggestion is supported by the metallicity variations which we find as functions of radial distance. 

The metallicity variation at larger vertical distances ($6.5<z\leq 9.5$ kpc) is small but monotonic. Three different vertical metallicity gradients could be detected: $d[M/H]/dz = -0.22(\pm0.03)$, $d[M/H]/dz = -0.38 (\pm0.06)$, and $d[M/H]/dz = -0.08 (\pm0.07)$ dex kpc$^{-1}$ for the intervals $z<3$, $3\leq z<5$, and $5\leq z<10$ kpc, respectively. Since our data cover the whole Galactic longitude interval, the resulting metallicity gradients can be interpreted as properties of the larger-scale Galaxy. The first gradient confirms the dissipational formation of the disc at short $z$--distances. The steeper gradient corresponds to the transition region between different population components of the Galaxy, and finally, the lowest value provides an adequate description of the inner-halo metallicity gradient.
\end{abstract}

\end{frontmatter}

\section{Introduction}
The metallicity distribution of G-type stars can be used for the interpretation of the formation and chemical evolution of the related Galactic components. For example, a metallicity gradient for a particular component of our Galaxy may indicate that this component formed by dissipational collapse, while the absence of a metallicity gradient may provide strong evidence of a merger and/or accretion origin from -- possibly numerous -- fragments, such as dwarf-type galaxies. The pioneering work on the dissipational collapse of the Galaxy is that of \citet{ELS62} who argued that the Galaxy collapsed in a free-fall time ($\sim2\times10^{8}$ yr). Now, we know that the Galaxy collapsed over many Gyr (e.g. \citet{YS79}; \citet{NBP85}; \citet{N86}; \citet{SF87}; \citet{CLL90}; \citet{NR91}; \citet{BSL95}). On the other hand, the merger or accretion scenario of galaxy formation was first advocated by \citet{SZ78}; for a survey of the more recent literature on the subject, we refer the reader to \citet{FBH02}.

Extant data comprise vertical metallicity gradients in the range $-0.40<d[M/H]/dz<-0.20$ dex kpc$^{-1}$ for relatively small distances from the Galactic plane, i.e., $z<4$ kpc (\citealt{Trefzger95, Karaali03, Du04, Ak07}), which support the picture of a dissipative collapse of the thin disc. For intermediate $z$-–distances, where the thick disc is dominant, the vertical metallicity gradient is low, $d[M/H]/dz=-0.07$ dex kpc$^{-1}$, and the radial gradient is only marginal, $-0.02\leq d[M/H]/dz\leq 0$ dex kpc$^{-1}$ (\citealt{RBK01}). We quote also the works of \citet{CB00} and \citet{Girard06} where vertical rotational velocity gradients were cited. In the first work, the authors give $\Delta <V_{\phi}>/\Delta |Z|=-52\pm6$ km s$^{-1}$ kpc$^{-1}$ for the halo stars whereas in the second work $\Delta <V_{\phi}>/\Delta|Z|=-30\pm3$ km s$^{-1}$ kpc$^{-1}$, for the stars dominated by the thick disc red giants. One must keep in mind that the samples assumed for a specific population, such as thin and thick discs or halo, may be contaminated by other objects. The work of \citet{Gilmore02} where higher mean rotational velocities than expected were predicted is a good example. Contrary to the expectations of the authors, the data of stars a few kpc from the Galactic plane which were assumed as candidates of the thick disc showed the mean rotation velocity $<V>\sim180$ km s$^{-1}$ instead of the $<V>\sim100$ km s$^{-1}$ one. The origin of most of the stars in the sample was a disrupted  satellite. Another important point in treating the metallicity or velocity gradient of a specific population is the effect of the metal-week component of the thick disc. If the separation of stars into different populations would be based mainly on their metallicities, the metal-week tail of the thick disc may cause some complications.           

In our recent papers, Galactic model parameters for 36 high-latitude fields have been estimated using photometric data from the Two Micron Sky Survey ({\em 2MASS\/}) and from the Sloan Digital Sky Survey ({\em SDSS\/}). For the thin and the thick discs, useful data have been available in both photometric surveys (\citealt{Cabrera-Lavers07, Bilir07}); however, for the halo only {\em SDSS} data could be used (\citealt{Bilir07}). We determined the variations of both the thick-disc scaleheight and the axial ratio of the halo as functions of Galactic longitude, and we noticed that it is most likely this variation which gives rise to the rather widely differing numerical values obtained for these parameters by different researchers.

Our work was based on de-reddened apparent $u$, $g$, $r$, $i$, $z$ magnitudes, estimated absolute magnitudes, and distances for 2 164 680 stars in a total area of 831 deg$^{2}$. In the present paper, we shall derive metal abundances for a subsample of these stars and combine them with the distances already estimated in our previous paper, in order to investigate the large-scale metallicity distribution, including vertical and radial metallicity gradients, and their possible dependence(s) on Galactic longitude.

\citet{Ak07} showed that in the anti-centre direction, the metallicity gradients corresponding to the transition region from the thick disc to the halo ($3<z\leq5$ kpc) are significantly different at the two different Galactic latitudes $b=+45^{o}$ and $b=-45^{o}$, respectively.
In the present study, the variation of metal-abundances and metallicity gradients with Galactic longitude, a trace of the triaxiality of the halo
will emerge as a particularly interesting result.

A brief description of the {\em SDSS} data and their reductions is given in Section 2. In section 3 we present the metallicity calibration for 36 fields and the longitude variation of the Galactic metal abundance distribution, including the vertical and radial metallicity gradients. These results are discussed in the final Section 4.  
      
\section{SDSS}
The {\em SDSS\/} is a large, international collaboration project set up to survey 10 000 square--degrees of sky in five optical passbands and to obtain spectra of one million galaxies, 100 000 quasars, and tens of thousands of Galactic stars. The data are being taken with a dedicated 2.5-m telescope located at Apache Point Observatory (APO), New Mexico. The telescope has two instruments: a CCD camera with 30 2048$\times$2048 CCDs in the focal plane and two 320 fiber double spectrographs. The imaging data are tied to a network of brighter astrometric standards (which would be saturated in the main imaging data) through a set of 22 smaller CCDs in the focal plane of the imaging camera. An 0.5-m telescope at APO has been used to tie the imaging data to brighter photometric standards. 

The {\em SDSS\/} obtains images almost simultaneously in five broad bands ($u$, $g$, $r$, $i$ and $z$)\footnote {Magnitudes in this paper are quoted in the $ugriz$ system to differentiate them from the former $u^{'}g^{'}r^{'}i^{'}z^{'}$ system.} centered at 3551, 4686, 6166, 7480 and 8932 $\AA$, respectively (\citealt{Fukugita96}). The imaging data are automatically processed through a series of software pipelines which find and measure objects and provide photometric and astrometric calibrations to produce a catalogue of objects with calibrated magnitudes, positions and structure information. The photometric pipeline (\citealt{Lupton01}) detects the objects, matches the data from the five filters, and measures instrumental fluxes, positions, and shape parameters (which allows the classification of objects as ``point source'', -compatible with the point spread function-, or ``extended''). The photometric calibration is accurate to roughly 2 per cent rms in the $g$, $r$ and $i$ bands, and 3 per cent in $u$ and $z$, as determined by the constancy of stellar population colours (\citealt{Ivezic04,Blanton05}), while the astrometric calibration precision is better than 0.1 arcsec rms per coordinate (\citealt{Pier03}). The Data Release 5 (DR5) imaging catalogue covers 8000 deg$^{2}$ (\citealt{Adelman07}) with a detection repeatability complete at a 95 per cent level for point sources brighter than the limiting apparent magnitudes of 22.0, 22.2, 22.2, 21.3 and 20.5 for $u$, $g$, $r$, $i$ and $z$, respectively. The data are saturated at about 14 mag in $g$, $r$ and $i$ and about 12 mag in $u$ and $z$. 

\subsection{Data and reductions}
The data used in this work were taken from the {\em SDSS\/} (DR5) WEB server\footnote{http://www.sdss.org/dr5/access/index.html} for 36 high-latitude fields ($60^\circ\leq b \leq65^\circ$) covering different Galactic longitude intervals throughout the full circle ($0^\circ< l \leq360^\circ$). From the 2 164 680 stars with {\em SDSS\/} magnitudes $u$, $g$, $r$, $i$, and $z$ used in our previous paper (\citealt{Bilir07}), we identified G-type stars by transformation of the Neil Reid's WEB page \footnote{http://www-int.stsci.edu/$\sim$inr/intrins.html} $(U-B)_{0}$ and $(B-V)_{0}$ colours for the G spectral type dwarfs to the $(u-g)_{0}$ and $(g-r)_{0}$ colours for this work. The absolute magnitude range of these stars, 258 268 in number in a total area of 831 deg$^{2}$, is $5 <M(g)\leq6$. The range of apparent $g$ magnitudes is $15<g_{0}\leq22$, as in our paper just cited. While the de-reddening of the apparent magnitudes, the determination of absolute magnitudes and the estimation of the distances relative to the Sun as well as the $z$--distances from the Galactic plane are explained in \citet{Bilir07}, derivation of the metal--abundances is described in the following Section 3.

\section{Metallicity}
The metallicities of the sample stars were evaluated using the following equation of \citet{KBT05}:
\begin{eqnarray}
[M/H]= 0.10-3.54\delta_{0.43}-39.63\delta_{0.43}^{2}+63.51\delta_{0.43}^{3}.
\end{eqnarray}

\begin{figure}
\begin{center}
\includegraphics[scale=0.4, angle=0]{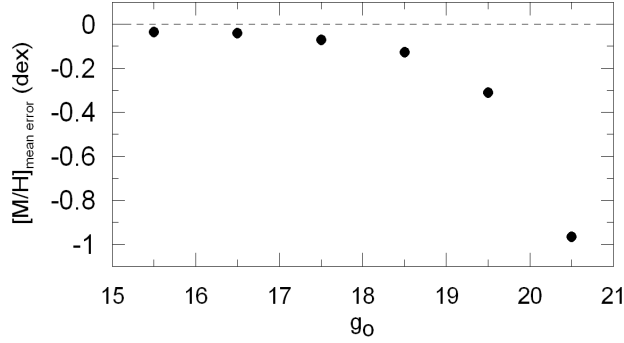}
\caption[]{Mean error versus $g_{0}$ apparent magnitude for stars with six apparent magnitude intervals, (15, 16], (16, 17], (17, 18], (18, 19], (19, 20] and (20, 21]. The mean error for $15<g_{0}\leq18$ where thick disc stars are dominant is less than $\pm$0.1 dex.}
\label{MH-error} 
\end{center}
\end{figure}

This equation was calibrated for the main sequence stars with $0.12<(g-r)_{0}\leq0.95$ which covers the $(g-r)_{0}$ colour indices of our sample. Here, $\delta_{0.43}$ is the normalized UV-excess in {\em SDSS\/} photometry corresponding to $\delta_{0.6}$ in the {\em UBV\/} photometry. \citet{KBT05} give the range of the metallicity as $-2.76\leq[M/H]\leq0.2$ dex, corresponding to $0<\delta_{0.43}\leq0.33$. Limitations of the photometric technique are unavoidable, especially for the metallicities of faint star. However, the thick disc stars occupy the intermediate apparent magnitude interval, i.e. $15<g_{0}\leq18$, (\citealt{Chen01}) where the mean error for $[M/H]$ is less than $\pm$0.1 dex (Fig. \ref{MH-error}). We emphasize that the parameters based on the determination of the metallicity are limited by the accuracy of metallicities estimates based on photometry alone, and that they should be supplemented with investigations of the soon to emerge spectroscopic determinations of parameters for the {\em SDSS\/} samples.    

The parameter range $0<\delta_{0.43}\leq0.33$ covers the metallicity interval $-3<[M/H]\leq0.2$ dex. 

The metallicity distribution for each field is given by a mean metal-abundance value determined for each of the following distance intervals (in kpc): (1.5,2], (2,2.5], (2.5,3], (3,4], (4,5], (5,6], (6,7], (7,8], (8,9], (9,12], (12,15]. These intervals and the corresponding centroid distances ($r^{*}$) are the same as in \citet{Ak07}; however, the projection of $r^{*}$ onto the vertical direction, i.e. $z=r^{*}\sin(b)$, is different due to the higher Galactic latitudes studied in this work.

Results are presented in Table 1. Notice that the metallicity distributions in the distance intervals $3<r\leq4$, $4<r\leq5$, $5<r\leq6$, and $6<r\leq7$ kpc are rather flat. Hence, median values could be used in doing the statistics for these intervals, whereas modal means were preferred for the other distance intervals which showed a peak.

\begin{table*}
{\tiny
\center
\caption{The metallicity distribution as a function of distance from the Galactic plane, for 36 high latitude ($60^\circ\leq b \leq65^\circ$) fields at different Galactic longitudes ($0^\circ < l \leq360^\circ$). Uncertainties (given in parentheses) refer to the last quoted digits. Distances in kpc.}
\begin{tabular}{cccccccccccc}
\hline
$l/z$& 1.07     &  1.58  &     2.02 &      2.46 &      3.17 &      4.04 &      4.92 &      5.80 &      6.68 &      7.56 &      9.50\\
\hline
0   & -0.38(09) & -0.57(06) & -0.68(06) & -0.57(06) & -1.21(08) & -1.61(11) & -1.70(11) & -1.78(11) & -2.11(09) & -2.33(12) & -2.28(07)\\
10  & -0.22(13) & -0.50(09) & -0.65(06) & -0.78(08) & -1.14(08) & -1.60(11) & -1.80(11) & -1.75(11) & -2.29(12) & -2.36(08) & -2.31(09)\\
20  & -0.53(04) & -0.63(07) & -0.63(06) & -0.73(04) & -1.22(07) & -1.64(11) & -1.79(11) & -1.82(11) & -2.19(11) & -2.25(14) & -2.41(10)\\
30  & -0.50(07) & -0.66(05) & -0.79(04) & -0.90(06) & -1.24(07) & -1.61(11) & -1.85(11) & -1.85(15) & -2.18(11) & -2.16(17) & -2.34(08)\\
40  & -0.55(04) & -0.60(04) & -0.73(04) & -0.75(07) & -1.19(08) & -1.58(11) & -1.85(11) & -1.93(11) & -2.28(05) & -2.38(22) & -2.27(08)\\
50  & -0.50(05) & -0.67(05) & -0.82(07) & -0.85(07) & -1.35(08) & -1.70(11) & -2.00(11) & -2.07(11) & -2.35(02) & -2.36(11) & -2.43(13)\\
60  & -0.50(05) & -0.69(04) & -0.72(05) & -0.82(06) & -1.18(11) & -1.68(11) & -1.89(11) & -2.07(15) & -2.28(10) & -2.28(14) & -2.29(11)\\
70  & -0.50(04) & -0.58(04) & -0.65(05) & -0.76(06) & -1.22(11) & -1.66(11) & -1.87(15) & -1.97(15) & -2.35(10) & -2.29(10) & -2.38(09)\\
80  & -0.47(08) & -0.62(03) & -0.62(06) & -0.78(20) & -1.29(11) & -1.67(15) & -1.85(15) & -1.93(15) & -2.35(06) & -2.47(11) & -2.31(13)\\
90  & -0.58(05) & -0.66(04) & -0.88(04) & -0.82(09) & -1.24(11) & -1.57(15) & -1.92(19) & -1.94(19) & -2.27(06) & -2.32(07) & -2.25(11)\\
100 & -0.55(04) & -0.60(07) & -0.74(09) & -0.95(11) & -1.25(11) & -1.67(15) & -1.88(15) & -2.03(19) & -2.26(05) & -2.38(13) & -2.27(11)\\
110 & -0.54(06) & -0.60(04) & -0.78(05) & -0.86(13) & -1.16(11) & -1.64(15) & -1.99(19) & -2.03(19) & -2.40(12) & -2.46(17) & -2.22(12)\\
120 & -0.29(12) & -0.43(13) & -0.71(11) & -0.96(07) & -1.20(15) & -1.50(15) & -1.78(19) & -1.98(19) & -2.37(21) & -2.29(12) & -2.08(10)\\
130 & -0.45(08) & -0.47(09) & -0.40(13) & -0.71(15) & -1.18(15) & -1.62(19) & -1.89(19) & -1.90(23) & -2.40(30) & -2.26(10) & -2.24(13)\\
140 & -0.42(09) & -0.70(05) & -0.73(08) & -1.09(14) & -1.23(15) & -1.73(19) & -1.89(19) & -1.95(19) & -2.45(10) & -2.39(20) & -2.30(14)\\
150 & -0.50(08) & -0.55(06) & -0.74(06) & -0.76(07) & -1.30(15) & -1.73(19) & -1.94(19) & -2.03(19) & -2.48(13) & -2.25(08) & -2.31(17)\\
160 & -0.32(10) & -0.51(04) & -0.38(09) & -0.50(10) & -1.22(15) & -1.59(19) & -1.93(19) & -1.92(19) & -2.30(15) & -2.42(07) & -2.43(39)\\
170 & -0.47(06) & -0.60(04) & -0.72(07) & -0.74(13) & -1.16(15) & -1.73(19) & -1.95(19) & -1.99(19) & -2.33(15) & -2.20(27) & -2.34(12)\\
180 & -0.44(08) & -0.57(06) & -0.73(13) & -0.69(08) & -1.22(15) & -1.71(15) & -1.96(19) & -1.99(19) & -2.38(03) & -2.34(13) & -2.28(06)\\
190 & -0.36(07) & -0.46(09) & -0.56(10) & -0.96(05) & -1.13(15) & -1.61(19) & -1.84(19) & -1.88(19) & -2.47(08) & -2.42(11) & -2.30(11)\\
200 & -0.30(13) & -0.45(07) & -0.52(06) & -0.77(04) & -1.20(15) & -1.64(15) & -1.91(19) & -2.02(19) & -2.24(09) & -2.42(05) & -2.32(11)\\
210 & -0.41(07) & -0.64(08) & -0.61(07) & -0.61(10) & -1.22(15) & -1.74(15) & -1.94(15) & -1.96(19) & -2.23(10) & -2.32(12) & -2.37(07)\\
220 & -0.44(06) & -0.66(06) & -0.64(04) & -0.74(12) & -1.19(15) & -1.62(15) & -1.89(15) & -1.85(15) & -2.57(47) & -2.32(11) & -2.33(09)\\
230 & -0.32(14) & -0.62(04) & -0.54(06) & -0.59(16) & -1.12(11) & -1.53(15) & -1.87(19) & -1.82(19) & -2.34(18) & -2.34(20) & -2.36(18)\\
240 & -0.50(07) & -0.54(05) & -0.68(08) & -0.64(06) & -1.18(11) & -1.55(15) & -1.86(15) & -1.84(19) & -2.34(09) & -2.30(09) & -2.39(16)\\
250 & -0.52(07) & -0.67(03) & -0.74(04) & -0.69(21) & -1.32(11) & -1.74(15) & -1.97(15) & -1.98(19) & -2.31(05) & -2.37(07) & -2.26(09)\\
260 & -0.49(05) & -0.69(06) & -0.75(05) & -0.86(12) & -1.16(11) & -1.69(15) & -1.95(15) & -1.78(19) & -2.36(14) & -2.15(13) & -2.34(09)\\
270 & -0.47(05) & -0.55(09) & -0.61(05) & -0.78(08) & -1.18(11) & -1.44(11) & -1.87(15) & -2.11(15) & -2.45(10) & -2.43(09) & -2.28(10)\\
280 & -0.43(04) & -0.63(04) & -0.73(08) & -0.65(08) & -1.26(11) & -1.62(11) & -1.86(11) & -1.95(15) & -2.19(06) & -2.36(12) & -2.37(12)\\
290 & -0.42(04) & -0.66(04) & -0.72(05) & -0.81(07) & -1.11(11) & -1.70(11) & -1.81(15) & -1.98(15) & -2.44(15) & -2.28(05) & -2.42(11)\\
300 & -0.51(03) & -0.58(06) & -0.73(09) & -0.92(14) & -1.20(08) & -1.58(11) & -1.84(11) & -2.05(11) & -2.32(03) & -2.69(51) & -2.32(06)\\
310 & -0.35(06) & -0.63(04) & -0.68(04) & -0.85(05) & -1.23(08) & -1.53(11) & -1.80(11) & -1.91(15) & -2.36(16) & -2.32(11) & -2.37(13)\\
320 & -0.45(06) & -0.64(04) & -0.74(06) & -0.82(12) & -1.23(08) & -1.52(11) & -1.75(11) & -1.87(11) & -2.35(08) & -2.43(07) & -2.28(08)\\
330 & -0.40(05) & -0.57(06) & -0.68(06) & -0.78(05) & -1.16(07) & -1.54(11) & -1.62(11) & -1.68(11) & -2.46(10) & -2.42(05) & -2.43(18)\\
340 & -0.25(11) & -0.58(06) & -0.67(05) & -0.74(06) & -1.19(07) & -1.51(11) & -1.77(11) & -1.77(11) & -2.45(15) & -2.43(08) & -2.25(10)\\
350 & -0.50(04) & -0.62(04) & -0.65(05) & -0.73(06) & -1.28(07) & -1.63(11) & -1.80(11) & -1.88(11) & -2.50(10) & -2.27(27) & -2.20(10)\\
\hline
\end{tabular}  
}
\end{table*}

\subsection{Metallicity variation with Galactic longitude}

\begin{figure}
\begin{center}
\includegraphics[scale=0.4, angle=0]{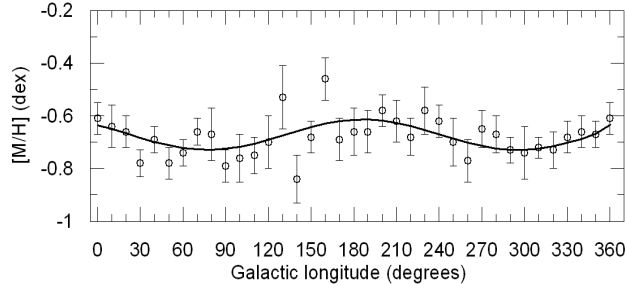}
\caption[]{Mean metallicity distribution as a function of Galactic longitude, for vertical distances $1<z<2.5$ kpc.}
\label{l-mh-tk} 
\end{center}
\end{figure}

The metallicity distributions for both (relatively) short and large $z$ distances show systematic fluctuations. For example, the distribution for $z=2$ kpc and the combined distribution for $1<z<2.5$ kpc have maxima at intermediate longitudes, while minima exist at small and large longitudes (Fig. \ref{l-mh-tk}). These $z$ distances correspond to the region where the discs -- both the thin and thick discs -- dominate the stellar distributions of the Galaxy. If compared to the typical error bars, the amplitude of the fluctuation is significant but only moderate, i.e. $\Delta[M/H]\sim0.2$ dex. However, if it is real, this feature should be related with the structure of the Galactic disc, probably with its warp and flare (see Section 4). For completeness, the longitude variation of Fig. \ref{l-mh-tk} has been fitted by the following equation (solid curve): 

\begin{eqnarray}
[M/H]=-0.67-0.06\sin[2\pi(0.0046l+0.3940)].
\end{eqnarray}

\begin{figure}
\begin{center}
\includegraphics[scale=0.41, angle=0]{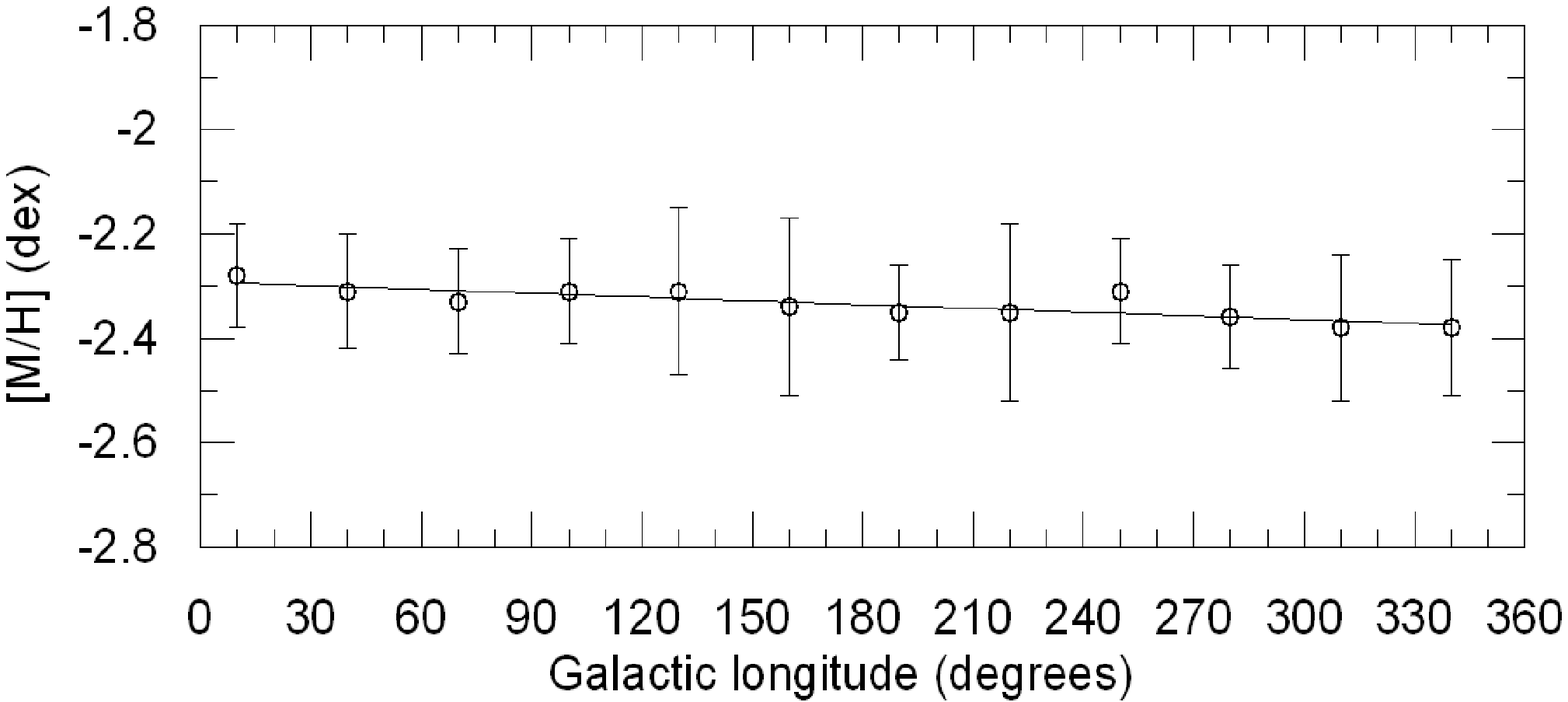}
\caption[]{Mean metallicity distribution as a function of Galactic longitude, for vertical distances $6.5<z\leq9.5$ kpc.}
\label{l-mh-hh} 
\end{center}
\end{figure}

The metallicity distribution for the large $z$ distances is even more interesting. Combining the distributions for $6.5<z\leq9.5$ kpc reveals a clear longitudinal metallicity distribution (Fig. \ref{l-mh-hh}). The metal abundance decreases monotonically from -2.28 to -2.38 dex when one goes from the Galactic longitude 10$^\circ$ to 340$^\circ$. However, this metallicity distribution can also be considered flat within the errors. The interesting is that the cited longitudes, 10$^\circ$ and 340$^\circ$, are close to the direction of the Galactic centre, where one expects the same metallicity. The distance interval $6.5<z\leq9.5$ kpc covers the halo component. Hence, this unexpected finding should be explained by the structure of the halo. The equation fitted to the metallicity distribution as follows:
\begin{eqnarray}
[M/H]=-0.00024l-2.292.
\end{eqnarray}
   
\subsection{Vertical and radial metallicity gradients}
\begin{figure}
\begin{center}
\includegraphics[angle=0, width=140mm, height=99mm]{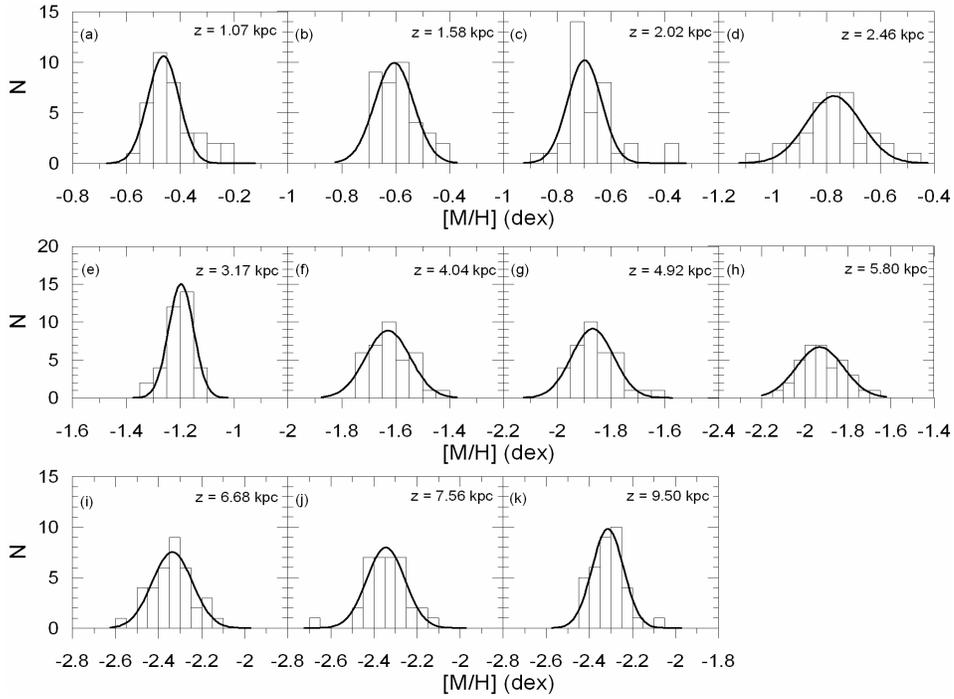} 
\caption[]{Metallicity histograms for the mean metallicities of 36 fields.}
\label{N-metal} 
\end{center}
\end{figure}

\begin{figure}
\begin{center}
\includegraphics[scale=0.7, angle=0]{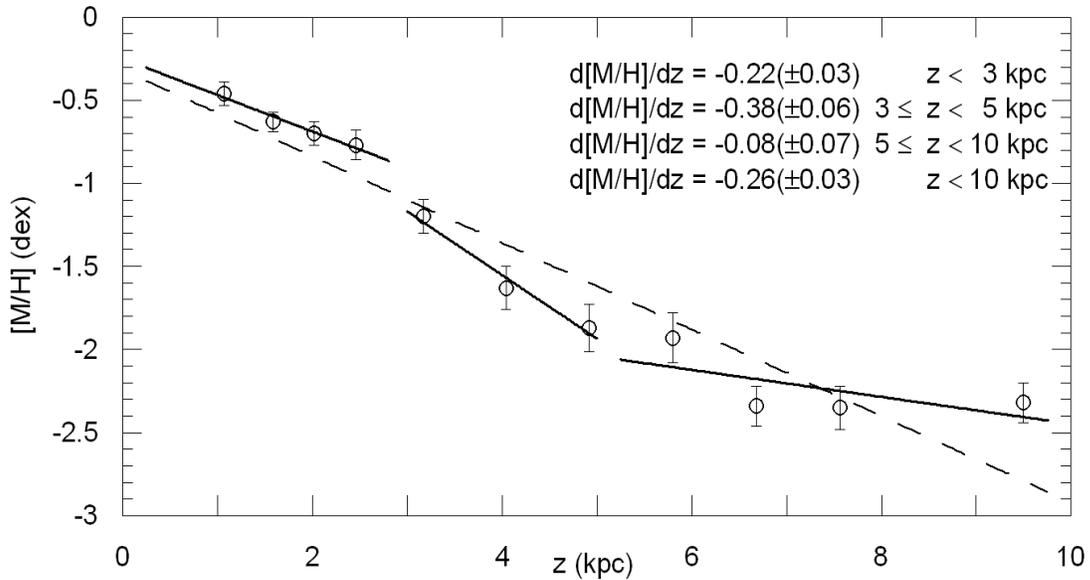}
\caption[]{Mean metallicity for 36 fields as a function of vertical distance $z$.}
\label{z-mh} 
\end{center}
\end{figure}

Figure \ref{N-metal} displays the histograms of the metallicities given in Table 1 for the 36 fields in each of 11 $z$ distance intervals covering the range $1<z\leq9.5$ kpc. Each histogram has been fitted by a Gaussian curve with modal mean adopted as equal to the mean metallicity of 36 fields at the corresponding $z$ distance from the Galactic plane (Table 2). As expected, the mean metallicities decrease with increasing $z$ distances, indicating a probable vertical metallicity gradient. We plotted the same data in Fig. \ref{z-mh} for investigating this process in detail. One can see three different trends in the figure: 1) for relatively short $z$ distances, i.e. $z<3$ kpc, the variation of $[M/H]$ is rather smooth, 2) for intermediate $z$ distances, i.e. $3\leq z<5$ kpc, the variation is steeper but still smooth, and finally 3) for $5\leq z<10$ kpc, the variation is flat, but fluctuations and error bars are larger. The metallicity gradient describing the first trend, $d[M/H]/dz=-0.22$ dex kpc$^{-1}$, is in agreement with the canonical metallicity gradients for the same $z$ distances, and is the likely signature of this Galactic region's formation by a dissipative collapse. The description of the second trend in terms of a metallicity gradient, $d[M/H]/dz=-0.38$ dex kpc$^{-1}$, corresponds to the average metallicity difference between the two population components involved. Finally, the third trend, $d[M/H]/dz=-0.08$ dex kpc$^{-1}$, is the low metallicity gradient of the inner spheroid.         

\begin{table}
\center
\caption{Mean metallicities for different distance intervals calculated from the metallicity distributions of 36 fields. Symbols: $r$: distance from the Sun; $r^{*}$: centroid distance corresponding to the interval $r_{1}-–r_{2}$; $z$: projection of $r^{*}$ onto the vertical direction; $[M/H]$: mean metallicity.}
\begin{tabular}{cccc}
\hline
$r_{1}-–r_{2}$ & $r^{*}$ & $z$ & $[M/H]$ \\
(kpc)         & (kpc)   &  (kpc)  & (dex)   \\
\hline
 0.5-–1.5 &       1.20 &       1.07 & -0.46(07) \\
 1.5-–2.0 &       1.78 &       1.58 & -0.63(06) \\
 2.0--2.5 &       2.28 &       2.02 & -0.70(07) \\
 2.5--3.0 &       2.77 &       2.46 & -0.77(09) \\
 3.0--4.0 &       3.57 &       3.17 & -1.20(10) \\
 4.0--5.0 &       4.56 &       4.04 & -1.63(13) \\
 5.0--6.0 &       5.54 &       4.92 & -1.87(14) \\
 6.0--7.0 &       6.54 &       5.80 & -1.93(15) \\
 7.0--8.0 &       7.53 &       6.68 & -2.34(12) \\
 8.0--9.0 &       8.53 &       7.56 & -2.35(13) \\
9.0-–12.0 &      10.71 &       9.50 & -2.32(12) \\
\hline
\end{tabular}  
\end{table}

\begin{figure}
\begin{center}
\includegraphics[scale=0.6, angle=0]{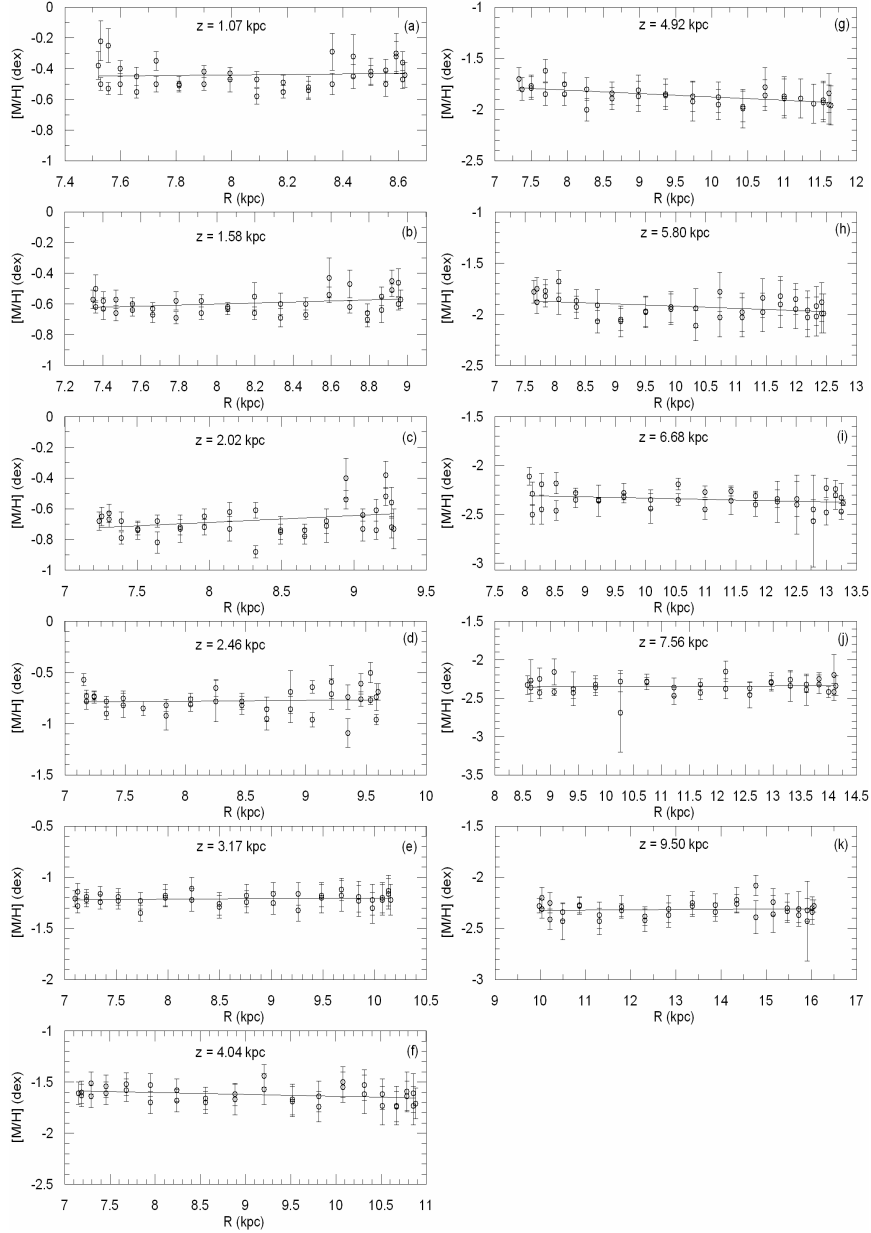}
\caption[]{Variation of the metallicities for 36 fields with radial distance $R$.}
\label{DMH-R} 
\end{center}
\end{figure}

We investigated the radial metallicity gradient for 36 fields as follows. It will be seen immediately that the radial metallicity gradient described and used in this work is different from the usual definition: although we defined metallicities for 36 fields for a specific $z$ distance from the Galactic plane, the distances of the centers of these fields to the Galactic centre are, of course, different due to their different Galactic longitudes, $0^\circ<l \leq360^\circ$. The range of radial distances thus defined is $7<R<16$ kpc. Metallicities for 36 fields in Table 1 as functions of radial distance $R$ are given in 11 panels in Fig. \ref{DMH-R}. From the definition, short radial distances correspond to fields in the general direction toward the centre of the Galaxy, whereas fields in the general direction toward the anti-centre have larger radial distances. 

Different trends can be observed in different panels. The most conspicuous features in panel (a) are the following: the fields in both the centre and anti-centre directions are relatively richer in metallicity than the fields in other directions, and their corresponding error bars are larger; furthermore, the metallicity differences between the centre and anti-centre fields follow an axisymmetric pattern.

The relative metal overabundances in the anti-centre fields (i.e., for large radial distances) are even larger in panels (b) and (c). The metallicity variations in panels (d) and (e) are almost zero, and finally, the metallicity trend changes sign for higher $z$ distances, i.e., $4<z<7$ kpc (panels f-i). In short: fields in the centre direction of the Galaxy appear to be overabundant in metallicities relative to the other fields. Although this metallicity excess is relatively low, $\Delta[M/H]\sim0.15$ dex, it appears to be real because the present statistical analysis shows the metallicity trend behind it to be significant. 

For completeness, we should also mention that the last two panels, $z=7.56$ and $z=9.50$ kpc, show zero variation in metallicity. 

\begin{figure}
\begin{center}
\includegraphics[scale=0.5, angle=0]{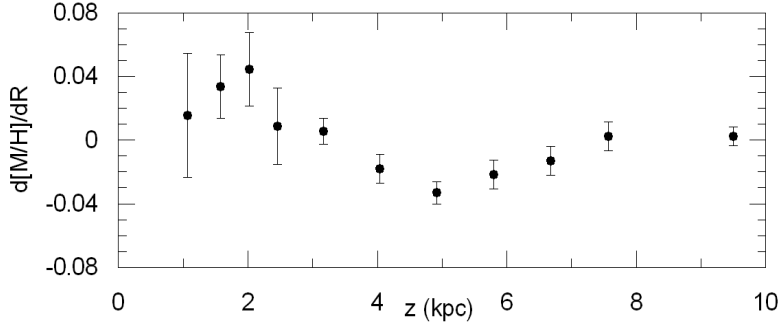}
\caption[]{Radial metallicity gradient versus vertical distance $z$.}
\label{dmhdr-z} 
\end{center}
\end{figure}

We fitted linear equations (solid lines) to the metallicity distributions in Fig. \ref{DMH-R}, and we plotted their slopes versus $z$ distances in Fig. \ref{dmhdr-z}. Clearly, the metallicity variation exhibits different trends at different distances from the Galactic plane. The minimum at $z\sim5$ kpc is the most conspicuous feature in this diagram, meaning that the slope of the radial metallicity gradient is significantly different at shorter and larger $z$ distances. It thus seems that the Galactic components (thin and thick discs, and halo) do not have homogeneous structures.

\section{Summary and Conlusion}

We evaluated metallicities for 36 high latitude ($60^\circ \leq b \leq 65^\circ$) fields with Galactic longitudes $0^\circ < l \leq 360^\circ$, and we investigated their variations with different parameters. 

The metallicities are longitude dependent. For relatively short $z$ distances from the Galactic plane, $z<2.5$ kpc, one can observe two minima at small and large Galactic longitudes ($l=80^\circ$ and $l=300^\circ$), respectively, and a maximum at $l=190^\circ$ (Fig. \ref{l-mh-tk}). For large distances, $6.5< z \leq 9.5$ kpc (Fig. \ref{l-mh-hh}), metallicity decreases monotonically with Galactic longitude. 

The metallicities of 36 fields were plotted as histograms, which were fitted by Gaussians giving a mean metallicity for each of 11 distinctive (representative) values of $z$ distance in the range $1<z\leq9.5$ kpc. The resulting mean metallicity as a function of $z$ reveals a trimodal behavior: (1) for $z<3$ kpc, there is a metallicity gradient of $d[M/H]/dz=-0.22$ dex kpc$^{-1}$ which is in agreement with the dissipational collapse picture of formation of the Galactic thin-disc component at relatively short $z$ distances; (2) a steeper gradient, $d[M/H]/dz=-0.38$ dex kpc$^{-1}$, covering the distance interval $3\leq z<5$ kpc which, although dominated by the thick disc, probably accentuates the metallicity differences between the coexisting thin and thick discs and the halo, rather than it provides a pure fossil record of thick-disc formation; (3)
the gradient $d[M/H]/dz=-0.08$ dex kpc$^{-1}$, for $5\leq z<10$ kpc, is a typical value for the inner spheroidal component.

These results, including the metallicity gradient $d[M/H]/dz=-0.26$ dex kpc$^{-1}$ for $z<10$ kpc, are all in fine agreement with the canonical literature (\citealt{Trefzger95, RBK01, Karaali03, Du04, Ak07}). By virtue of their systematic coverage of Galactic longitudes, they provide an important enhancement of our reconnaissance of Galactic structure which will be needed for an improved understanding of the formation of the Milky Way Galaxy. 

Conversion of vertical ($z$) distances to radial ($R$) distances also allowed a study of the metallicity variations as functions of $R$ over a range of almost 10 kpc. Again, clear (linear) systematic trends with different slopes can be identified behind the observed metallicity fluctuations, confirming results obtained by \citet{Allende06} from DR3 data for 22 700 F- and G-type stars. In particular, although these authors investigate a larger range in $R$, the metallicity structures that they find for the thick disc and the halo (their figures 13) are almost the same as those presented here for $z<2.5$ and $6.5<z\leq 9.5$ kpc.  

Our results are reminiscent of the flare effect of the disc. According to \citet{Lopez02}, a flare occuring in the outer disc produces an increase in the scaleheight as one moves radially outward. From their results we infer that the scaleheight of the thin disc sources is as large as 0.6 kpc for $R=10$ kpc, so it is plausible that a mixture of sources from the thin and thick discs are found in the outer disc at the lowest height regimes plotted in Fig. \ref {DMH-R} (panels a and b), producing higher metallicities than the expected from the contribution of the thick disc component alone.

The longitudinal dependence of the metallicity can be then explained by this process, since different longitudes correspond to different radial 
distances for a specific distance from the Galactic centre. The anticentre direction corresponds to the larger distances from the Galactic centre, thus is in this longitude range where the higher metallicities must be obtained due to the effect of the flare, as it is shown in Fig. \ref{l-mh-tk}. Also the longitudes where either the maximum or the minimum of the metallicity distribution are observed coincide with the directions of maximum warp amplitude in the Galactic thin disc (\citealt{Lopez02}). As this warp bends above and below the thin disc from its mean location, again mixing of stellar population from the thin and thick discs is expected at these longitudes: hence, observations of a variation in the mean metallicity are not surprising. Finally, even minor metallicity differences between fields at longitudes $l=10^\circ$ and $l=340^\circ$ (both in the Galactic centre direction), they may result from a triaxial structure of the halo -- or in other words: from the fact that the plane and the meridian of the disc are different from those of the halo. 

\section{Acknowledgments}
This work was supported by the Research Fund of the University of Istanbul. Project number: BYPF-1 12/31012007.

\end{document}